\documentclass[reprint,superscriptaddress,bibnotes,amsmath,amssymb,aps,prl,floatfix]{revtex4-1}

\usepackage{graphicx}% Include figure files
\usepackage{dcolumn}% Align table columns on decimal point
\usepackage{bm}% bold math
\usepackage{upgreek}
\usepackage{physics}
\usepackage{color}
\usepackage{amssymb}
\usepackage{hyperref}% add hypertext capabilities
\begin{document}

\preprint{APS/123-QED}

\title{Photon Statistics of Filtered Resonance Fluorescence}

\author{Catherine L. Phillips}
\affiliation{Department of Physics and Astronomy, University of Sheffield, Sheffield, S3 7RH, United Kingdom}
\author{Alistair J. Brash}
\email[Email:]{a.brash@sheffield.ac.uk}
\affiliation{Department of Physics and Astronomy, University of Sheffield, Sheffield, S3 7RH, United Kingdom}
\author{Dara P. S. McCutcheon}
\affiliation{Quantum Engineering Technology Labs, H. H. Wills
Physics Laboratory and Department of Electrical and Electronic Engineering, University of Bristol, Bristol, BS8 1FD, UK}
\author{Jake Iles-Smith}
\affiliation{Department of Physics and Astronomy, University of Sheffield, Sheffield, S3 7RH, United Kingdom}
\affiliation{Department of Physics and Astronomy, The University of Manchester, Oxford Road, Manchester, M13 9PL, United Kingdom}
\affiliation{Department of Electrical and Electronic Engineering, The University of Manchester, Sackville Street Building, Manchester, M1 3BB, United Kingdom}
\author{Edmund Clarke}
\affiliation{EPSRC National Epitaxy Facility, Department of Electronic and Electrical Engineering, University of Sheffield, Sheffield, S1 3JD, UK}
\author{Benjamin Royall}
\affiliation{Department of Physics and Astronomy, University of Sheffield, Sheffield, S3 7RH, United Kingdom}
\author{Maurice S. Skolnick}
\affiliation{Department of Physics and Astronomy, University of Sheffield, Sheffield, S3 7RH, United Kingdom}
\author{A. Mark Fox}
\affiliation{Department of Physics and Astronomy, University of Sheffield, Sheffield, S3 7RH, United Kingdom}
\author{Ahsan Nazir}\email[Email:]{ahsan.nazir@manchester.ac.uk}
\affiliation{Department of Physics and Astronomy, The University of Manchester, Oxford Road, Manchester, M13 9PL, United Kingdom}

\date{\today}

\begin{abstract}

Spectral filtering of resonance fluorescence is widely employed to improve single photon purity and indistinguishability by removing unwanted backgrounds. For filter bandwidths approaching the emitter linewidth, complex behaviour is predicted due to preferential transmission of components with differing photon statistics. We probe this regime using a Purcell-enhanced quantum dot in both weak and strong excitation limits, finding excellent agreement with an extended sensor theory model. By changing only the filter width, the photon statistics can be transformed between antibunched, bunched, or Poissonian. Our results verify that strong antibunching and a sub-natural linewidth cannot simultaneously be observed, providing new insight into the nature of coherent scattering.
\end{abstract}
\maketitle

Resonance fluorescence (RF) of two-level emitters (TLEs) is integral to numerous important proposals for optical quantum technologies such as single photon sources~\cite{Somaschi2016,PhysRevLett.116.213601,Liu2018}, spin-photon entanglement~\cite{DeGreve2012,Gao2012} and entanglement of remote spins~\cite{Delteil2015,PhysRevLett.119.010503}. The emitted spectrum is well-suited to these applications, exhibiting strong single-photon antibunching and, under appropriate excitation conditions, a dominant coherently scattered component with a sub-natural linewidth inherited from the laser coherence. Indeed, an ideal TLE in the limit of vanishing driving strength would exhibit both perfect antibunching and a coherent fraction approaching unity. Experimentally, studies have observed both strong antibunching and  high coherent fractions in separate measurements performed under identical conditions~\cite{Matthiesen2012,Proux2015,Bennett2016}. It is thus perhaps intuitive to assume that this coherent component must itself be antibunched. However this is not the case; by exploiting spectral filtering we demonstrate that, in accordance with theoretical predictions~\cite{dalibard1983correlation,Carreno2018}, antibunching requires interference between coherent and incoherent scattering and consequently cannot be observed simultaneously with a sub-natural linewidth.

In experimental quantum optics, spectral filtering around the zero phonon line (ZPL) of a TLE is widely employed to remove unwanted backgrounds from the driving laser~\cite{Somaschi2016,Liu2018}, other transitions~\cite{doi:10.1021/acs.nanolett.8b05132} or phonon sidebands~\cite{doi:10.1021/acs.nanolett.8b05132,PhysRevB.96.165306,PhysRevLett.122.173602}, improving the measured single photon purity and indistinguishability. Considering only indistinguishability, reducing the filter bandwidth always gives an improvement (at the cost of efficiency) as more background is removed~\cite{Iles-Smith2017}. However, as the filter bandwidth approaches the natural linewidth ($\gamma$) of the ZPL, theory predicts strongly modified photon statistics in both weak (coherent scattering) \cite{Carreno2018,Casalengua2019} and strong (Mollow triplet) \cite{Gonzalez-Tudela2013} driving regimes, an effect generally overlooked in experiments to date. Here, we experimentally verify these predictions, combining our results with a theoretical model to develop a thorough understanding of the complex photon statistics associated with spectrally filtered resonance fluorescence. These concepts are equally applicable to the broad assortment of atomic and atom-like TLEs used in current quantum optics research.

\begin{figure*}
	\centering
	\includegraphics{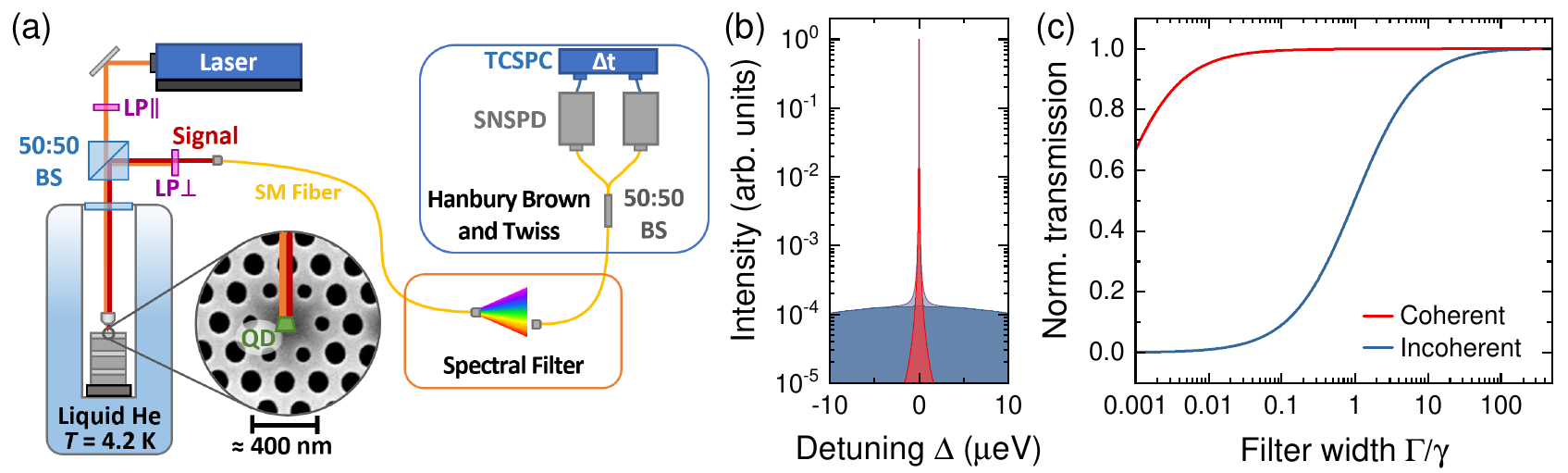}
	\caption{(a) Experimental set-up; LP - linear polariser, BS - beamsplitter, SM - single mode fiber, SNSPD - superconducting nanowire single photon detectors, TCSPC - time correlated single photon counter. (b) The calculated QD emission spectra (purple) under weak excitation comprises incoherent (blue) emission with 20 $\upmu$eV linewidth from spontaneous and stimulated emission and a narrow coherent (red) component that inherits the 10 neV linewidth of the CW laser; both components are modelled with a Lorentzian line shape. (c) Transmission coefficients of the coherent (red) and incoherent (blue) parts of the QD spectra through an ideal Lorentzian filter of width $\Gamma$. Changing $\Gamma$ strongly modifies the ratio of the two components.}
	\label{fig:setup_and_ratio}
\end{figure*}

The sample is studied in a liquid helium bath cryostat at 4.2~K and incorporates a self-assembled InGaAs quantum dot (QD) into an H1 photonic crystal cavity (PhCC) with coupled W1 waveguides (Fig$\ldotp$\ref{fig:setup_and_ratio}(a)). Resonant continuous wave (CW) laser excitation and collection of emission is made from directly above the cavity whilst laser back-scattering is rejected using a cross-polarisation technique. A p-i-n diode structure allows the QD neutral exciton to be electrically tuned. At the cavity resonance, a maximum Purcell factor of 43 shortens the QD's radiative lifetime ($T_{1}$) to 22.7~ps and results in lifetime-limited coherence~\cite{Liu2018}. Here, the QD is slightly detuned from the cavity, giving a Purcell factor of $\sim$30 and a broad natural linewidth ($\gamma$) of 20 $\upmu$eV.
This large $\gamma$ enables exploration of filter bandwidths ($\Gamma$) $\leq \gamma$ using a combination of diffraction grating and etalon filters (details in the Supplemental Material \cite{sup}).\nocite{PhysRevLett.109.183601,PhysRevLett.116.249902,PhysRevLett.105.177402,Nazir_2016,carmichael1,PhysRevB.95.201305,PhysRevLett.123.167403,PhysRevLett.104.017402}

For resonant CW excitation and a lifetime-limited emitter coherence time ($T_2 = 2T_1$), the weak excitation limit is defined as $\Omega_{R}^{2} < (\gamma^{2}/2)$ where $\Omega_{R}$ is the Rabi frequency and $\gamma = 1/T_1$~\cite{CohenTannoudji1998}. This is often termed the Resonant Rayleigh Scattering (RRS) or Heitler regime~\cite{Heitler1944,Nguyen2011,Matthiesen2012}. The RF spectrum in this regime includes contributions from both coherent RRS and an incoherent part originating from spontaneous and stimulated emission~\cite{Konthasinghe2012,Proux2015}. For coherent scattering, excitation and emission become a single coherent event where the elastically scattered photons inherit the laser coherence, leading to a \emph{sub-natural} linewidth~\cite{PhysRevLett.35.1426,Nguyen2011,Matthiesen2012,Konthasinghe2012,Bennett2016,Liu2018} that illustrates the long coherence times possible in this regime. Meanwhile, the \emph{natural linewidth} of the incoherent component is given by $\gamma=1/T_1$. Theory suggests that for weak excitation, interference between these different components is the origin of the observed photon antibunching~\cite{Carreno2018}. Owing to the discrepancy in linewidth between coherent and incoherent components, filtering with width $\Gamma < \gamma$ inevitably alters the ratio of the different components, modulating the interference between them and thus the observed photon statistics.

To explore this, Hanbury Brown and Twiss (HBT) measurements \cite{Brown1956} of the second-order correlation function ($\mathrm{g}^{(2)}(t)$) were performed in the weak CW driving regime. A value of $\mathrm{g}^{(2)}(0) < 1$ corresponds to antibunched emission whilst a value of $\mathrm{g}^{(2)}(0) = 1$ corresponds to the Poissonian statistics of a coherent source such as a laser. The QD is resonantly excited at laser energy $\hbar\omega_L$, inducing a Rabi frequency $\Omega_{R} = 0.5\:\gamma$. The emission is collected in cross-polarisation with signal-to-background ratio $> 100:1$ \cite{sup}. It then passes through a filter centred on the ZPL (details in Ref. \cite{sup}) before being split by a 50:50 fiber beamsplitter to a pair of superconducting nanowire single photon detectors (SNSPD) connected to a time-correlated single photon counting module (TCSPC), shown schematically in Fig.$\:$\ref{fig:setup_and_ratio}(a). The SNSPDs have a Gaussian instrument response function (IRF) with 37.5 $\pm$ 0.1 ps full-width half-maximum.

\begin{figure*}
    \centering
    \includegraphics{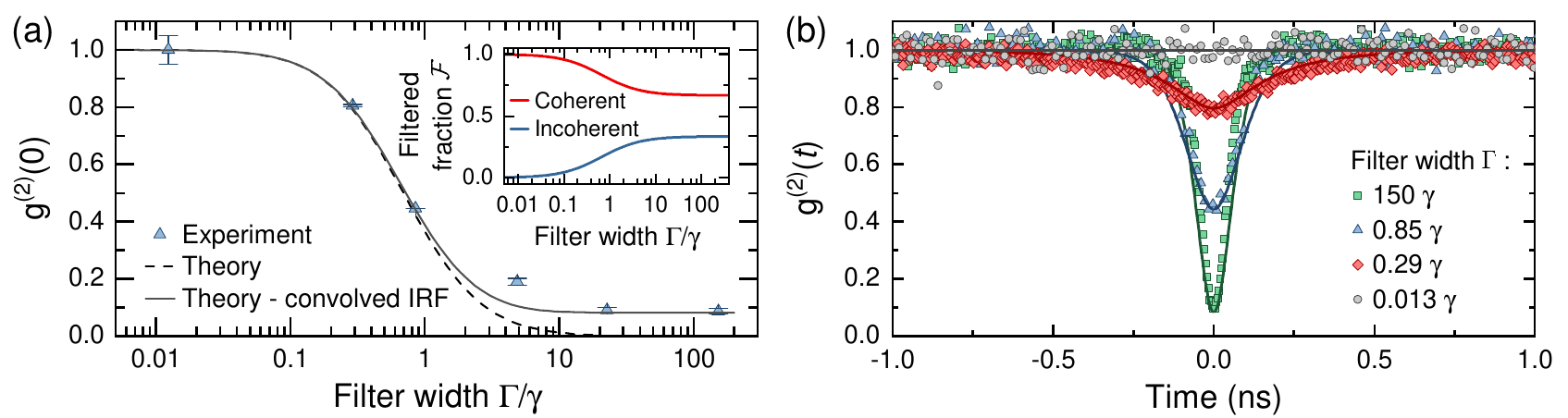}
    \caption{Filtered photon statistics under weak ($\Omega_{R} = 0.5\:\gamma$) driving: (a) $\mathrm{g}^{(2)}(0)$ measurements (blue triangles) of the exciton emission through different filter bandwidths ($\Gamma$), lines show the sensor theory prediction  with (solid) and without (dashed) convolution with the detector IRF. \textit{Inset:} Calculated fraction ($\mathcal{F}$) of the filtered spectrum originating from coherent (red) or incoherent (blue) scattering. (b) Full $\mathrm{g}^{(2)}(t)$ measurements from the same dataset exhibit both time broadening and a reduced antibunching dip at narrower filter bandwidths. The solid lines are a sensor theory calculation incorporating the detector IRF.}
    \label{fig:filterwidth_timebroadening_g2}
\end{figure*}

Fig$\ldotp$ \ref{fig:setup_and_ratio}(b) illustrates the theoretical total spectrum (purple) of the QD under these conditions (see Ref. \cite{sup} for corresponding experimental spectrum), comprising an incoherent peak with $\gamma = 20 ~\upmu$eV (blue) and a coherent peak with a linewidth of $\sim 10~\mathrm{neV}$ inherited from the laser (red). The area of the coherent peak relative to the total spectrum is the coherent fraction $\mathcal{F}_{CS}$ \cite{CohenTannoudji1998}:
\begin{equation}
    \mathcal{F}_{CS} = \frac{1}{1+2\Omega_{R}^{2}/\gamma^2},
    \label{eq:rrs_fraction}
\end{equation}
which gives $\mathcal{F}_{CS} = 2/3$ for $\Omega_{R} = 0.5\:\gamma$. The transmission coefficients of the coherent and incoherent parts through an ideal Lorentzian filter with bandwidth $\Gamma$ are plotted in Fig$\ldotp$ \ref{fig:setup_and_ratio}(c). As $\Gamma$ is reduced, the transmission of the incoherent component decreases much faster than the coherent component owing to the large ($2000 \times$) linewidth difference. Spectral filtering can thus manipulate this ratio up to a limiting case where narrow filtering removes the incoherent component almost entirely.

The variation of $\mathrm{g}^{(2)}(0)$ with $\Gamma$ is shown in Fig$\ldotp$ \ref{fig:filterwidth_timebroadening_g2}(a) for $\Omega_{R} = 0.5\:\gamma$. As expected for an unfiltered ideal TLE, strong antibunching is observed where the filter bandwidth exceeds the natural linewidth ($\Gamma > \gamma$). At $\Gamma = 150 \:\gamma$, $\mathrm{g}^{(2)}(0)=0.09 \pm 0.01$ limited only by the detector IRF ($= 1.14 \:\gamma^{-1}$). However, as the filter bandwidth becomes $\ll \gamma$, the antibunching is lost and $\mathrm{g}^{(2)}(0)$ tends towards $1$. The experiment agrees well with theoretical predictions (black lines in Fig$\ldotp$ \ref{fig:filterwidth_timebroadening_g2}(a)) derived using the sensor formalism \cite{PhysRevLett.109.183601,PhysRevLett.116.249902,sup} with (solid line) or without (dashed line) convolution with the detector IRF. 
%%%

The sensor theory is equivalent to the calculation of the correlation function from the RF electric field operators, with the sensor damping rates playing the role of the filter width. As previous works have shown~\cite{dalibard1983correlation}, the lowest order relevant term in the field contains a coherent and incoherent scattering contribution. These contributions destructively interfere to give zero when no filters are present, and this interference is partially or completely removed when filters are introduced.
%%%
It is interesting to note that when moving from $\Gamma = 150\:\gamma$ to $\Gamma = 23\:\gamma$, nearly the entire phonon sideband~\cite{PhysRevB.95.201305,PhysRevLett.123.167403,PhysRevLett.123.167402} is removed with no appreciable change in $\mathrm{g}^{(2)}(0)$. The nature of these measurements mean that electron-phonon interaction processes such as excitation-induced dephasing~\cite{PhysRevLett.104.017402} and phonon sideband emission~\cite{PhysRevB.95.201305,PhysRevLett.123.167403,PhysRevLett.123.167402} have negligible impact on $\mathrm{g}^{(2)}(t)$. Discussions of the sensor formalism, its extension to include laser background (see below) and phonon effects are given in the Supplemental Material~\cite{sup}. 

%Comparing 
From Figs$\ldotp$ \ref{fig:setup_and_ratio}(c) and \ref{fig:filterwidth_timebroadening_g2}(a) % illustrates 
we see that the loss of antibunching occurs in the regime ($0.1\:\gamma \lesssim \Gamma \lesssim 10 \:\gamma$) where the %reduced filter bandwidth 
filter removes almost the entire incoherent component. Indeed, the inset to Fig$\ldotp$ \ref{fig:filterwidth_timebroadening_g2} shows that in this region, the filtered coherent fraction approaches unity. This demonstrates that without both coherent and incoherent contributions, strong antibunching cannot be observed, indicating
%in accordance with theoretical predictions~\cite{dalibard1983correlation,Carreno2018} 
that the antibunching originates from interference between these components~\cite{dalibard1983correlation,Carreno2018}. 
We note that if it were possible to similarly remove only the coherent component, bunched statistics would be expected~\cite{dalibard1983correlation}.

Fig$\ldotp$ \ref{fig:filterwidth_timebroadening_g2}(b) shows some of the individual $\mathrm{g}^{(2)}(t)$ measurements from which Fig$\ldotp$ \ref{fig:filterwidth_timebroadening_g2}(a) is derived. As the filter bandwidth narrows, the central dip in $\mathrm{g}^{(2)}(t)$ broadens in width. This can be interpreted according to the uncertainty relation $\Delta E \Delta t > \frac{\hbar}{2}$ which implies that a narrower filter ($\Delta E$) inevitably increases the associated timing uncertainty of the photon. Considering that filtering with bandwidth $\Gamma$ is equivalent to a projective measurement of a photon linewidth $< \Gamma$~\cite{dalibard1983correlation,Carreno2018}, this illustrates that it is impossible to simultaneously observe both a sub-natural linewidth and strong antibunching from a TLE.

\begin{figure}
    \centering
    \includegraphics{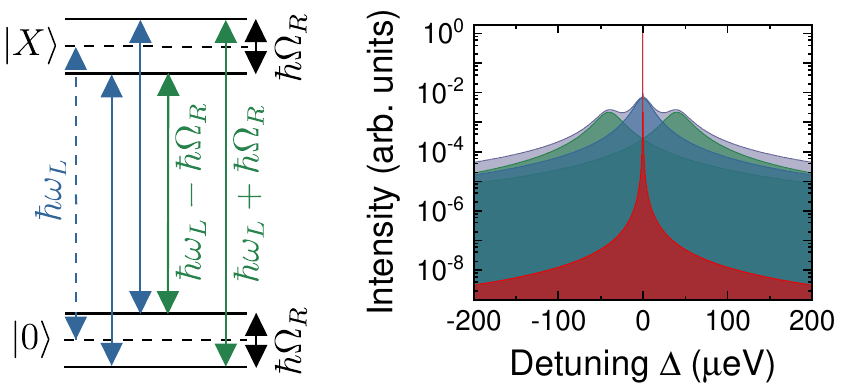}
    \caption{Theoretical spectrum at $\Omega_R = 2\:\gamma$. Strong driving splits the ground and excited states (dashed) by the Rabi energy ($\hbar\Omega_R$). Two of the four transitions (blue/green) are degenerate (blue), creating a Mollow triplet spectrum (purple) with a narrow coherent component also present (red).}
    \label{fig:Mollow}
\end{figure}

%Beyond the weak excitation limit, 
Looking now at the strong driving regime defined as $\Omega_{R}\gg \frac{1}{T_{2}}$, Fig.$\:$\ref{fig:Mollow} shows the resulting AC Stark effect transformation of the \textquotedblleft bare\textquotedblright$\:$states of the TLE into \textquotedblleft dressed\textquotedblright$\:$states split by the Rabi energy ($\hbar\Omega_{R}$). This splitting gives four possible transitions between upper and lower manifolds; as two of the transitions are degenerate, the result is the purple Mollow triplet spectrum shown in Fig.$\:$\ref{fig:Mollow} for $\Omega_R = 2\gamma$. The central (Rayleigh) peak is %often termed the Rayleigh peak (blue), 
flanked by two side (Mollow) peaks. 
The width of the individual peaks is governed by $\gamma$ \cite{ulhaq_detuning-dependent_2013,PhysRevLett.110.217401}. In addition to these incoherent peaks, a contribution from coherent scattering remains (red). As $\Omega_R$ increases, the Mollow splitting between side peaks increases whilst the coherent fraction decreases according to Eq.$\:$\ref{eq:rrs_fraction}.

\begin{figure*}
    \centering
    \includegraphics{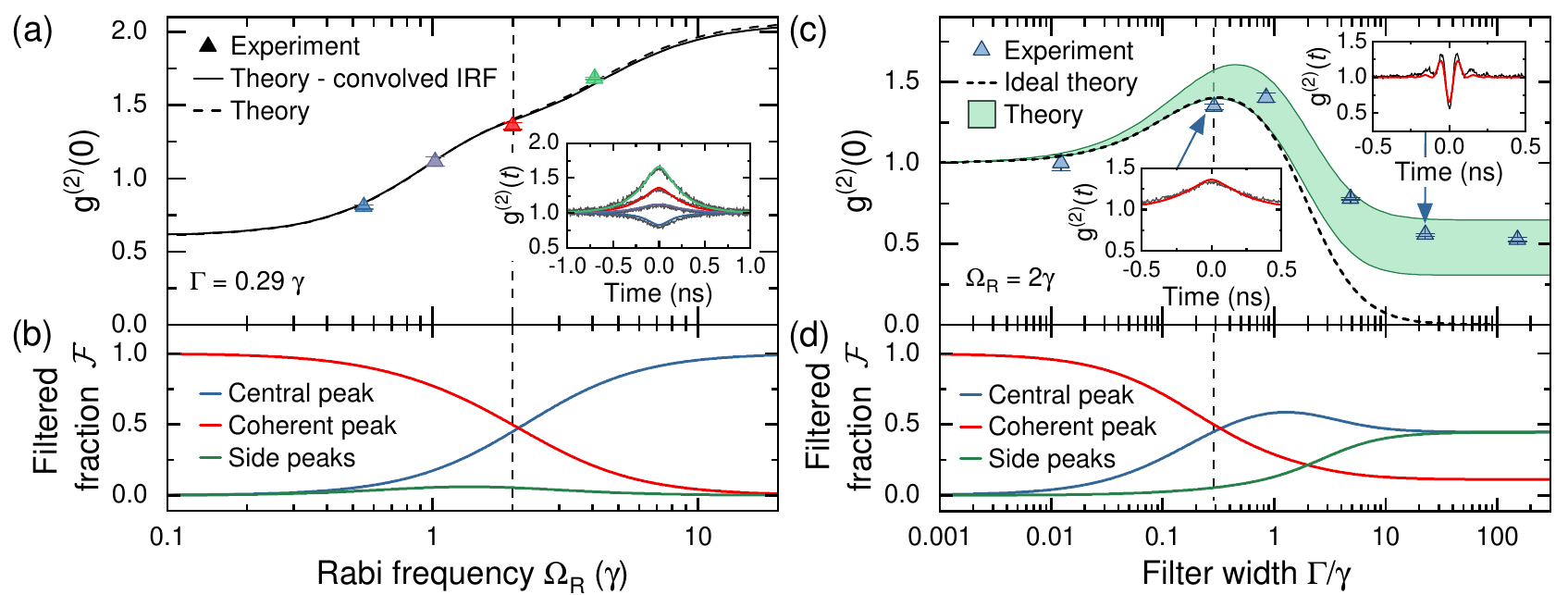}
    \caption{Filtered photon statistics under strong driving: (a) Measurements (triangles) of $\mathrm{g}^{(2)}(0)$ filtered at $\Gamma = 0.29\:\gamma$ transition from antibunching to bunching with increasing $\Omega_{R}$, lines show the sensor theory prediction with (solid) and without (dashed) detector IRF convolution. \textit{Inset:} Full $\mathrm{g}^{(2)}(t)$ measurements from same dataset; lines - model with fitted laser background level. (b) Calculated fraction ($\mathcal{F}$) of the filtered ($\Gamma = 0.29\;\gamma$) spectrum originating from each Mollow triplet component. (c) Measurements (triangles) of $\mathrm{g}^{(2)}(0)$ at $\Omega_{R} = 2\gamma$ through different filter widths ($\Gamma$); dashed line - theory prediction for ideal case, green region - confidence bounds of theory including IRF and laser backgrounds between 0 (lower) and $20~\%$ (upper) of the total signal. \textit{Insets:} Full $\mathrm{g}^{(2)}(t)$ measurements from the same dataset, Rabi oscillations are observed for $\Gamma > \gamma$. Lines - model with fitted laser background level. (d) Calculated fraction ($\mathcal{F}$) of the filtered spectrum originating from each component. Dashed lines (at $\Omega_R = 2\:\gamma$ and $\Gamma = 0.29\:\gamma$) are equivalent points for comparison between Figs. \ref{fig:highpower}~(a,b) and (c,d).
    }
    \label{fig:highpower}
\end{figure*}

Frequency-resolved studies of Mollow triplet photon correlations have revealed a rich assortment of physics. An unfiltered Mollow spectrum exhibits antibunching whilst isolating individual peaks results in $\mathrm{g}^{(2)}(0) = 1$ for the central Rayleigh peak and antibunching for the side peaks~\cite{Schrama1992,Ulhaq2012,Weiler2013}. Cross-correlation measurements between the Rayleigh peak and either side peak exhibit antibunching~\cite{Schrama1991} whilst a cross-correlation between side peaks exhibits bunching ($\mathrm{g}^{(2)}(0)>1$) \cite{Ulhaq2012,Schrama1992}. In addition, filtering half-way between the central and side peaks has revealed the existence of weak \textquotedblleft leapfrog\textquotedblright~two-photon transitions that exhibit strong bunching~\cite{Peiris2015,PhysRevB.98.165432}.

The aforementioned studies were performed with broad filtering ($\Gamma > \gamma$), aside from Ref. \cite{Peiris2015} where weak bunching ($\mathrm{g}^{(2)}(0) \sim 1.2$) was observed when the Rayleigh peak was filtered at $\Gamma \sim 0.25 \:\gamma$. Here, the large $\gamma$ of our sample facilitates thorough exploration of this regime. We begin by measuring $\mathrm{g}^{(2)}(0)$ as a function of $\Omega_R$, filtering centred on the Rayleigh peak with $\Gamma = 0.29 \:\gamma$. The results (Fig.$\:$\ref{fig:highpower}(a)) illustrate a surprising transition from antibunching to strong bunching with increasing $\Omega_{R}$.

To understand this result requires careful consideration of the relationship between the Rabi frequency $\Omega_R$ and the amplitude and filter transmission of the various components of the RF spectrum. The fraction ($\mathcal{F}$) of the filtered ($\Gamma = 0.29 \:\gamma$) spectrum arising from each Mollow triplet component is plotted against $\Omega_R$ in Fig.$\:$\ref{fig:highpower}(b). At small $\Omega_{R}$, Eq.$\:$\ref{eq:rrs_fraction} dictates a large coherent fraction. Thus, the behaviour in this region corresponds to Fig.$\:$\ref{fig:filterwidth_timebroadening_g2}(a); only weak antibunching is observed as the filter bandwidth $< \gamma$ removes the majority of the incoherent component. As $\Omega_{R}$ increases, the coherent fraction falls and the splitting of the Mollow triplet increases, reducing the transmission of the side peaks (green) through the filter. It is thus intuitive to expect a transition to the Poissonian statistics of the Rayleigh peak (blue)~\cite{Schrama1992,Ulhaq2012,Weiler2013} that now dominates the filtered spectrum.

However, in the limit $\Gamma < \gamma$, the additional effect of \textquotedblleft indistinguishability bunching\textquotedblright$\:$\cite{Gonzalez-Tudela2013} also becomes relevant. This phenomena originates in the quantum fluctuations of the light field~\cite{Armstrong:66,Knoll:86} and has been observed to lead to photon bunching when filtering at less than the natural linewidth of a light source, even for a classical input state such as a laser~\cite{neelen_spectral_1993}. In the case of the RF spectrum considered here, the filtering is narrow compared to the incoherent Rayleigh peak but still broad compared to the coherent component. As such, for larger $\Omega_R$ where side peak contributions are negligible, the filtered $\mathrm{g}^{(2)}(0)$ of Fig.$\:$\ref{fig:highpower}(a) is determined by competition between the Poissonian statistics of the coherent part (see Fig.$\:$\ref{fig:filterwidth_timebroadening_g2}(a)) and bunching originating from the narrowly filtered incoherent part. Therefore, as $\Omega_R$ increases, the decreasing coherent fraction allows the indistinguishability bunching effect to dominate, leading to the strong bunching observed for large $\Omega_R$ in Fig.$\:$\ref{fig:highpower}(a).

Our theoretical model (solid line in Fig.$\:$\ref{fig:highpower}(a)) reproduces well the experimental results and predicts a maximum bunching of $\mathrm{g}^{(2)}(0) \sim 2.1$ for these parameters. Experimentally, measurements cannot accurately be made at $\Omega_R > 4 \gamma$ owing to increasing laser background. We note that theoretical studies \cite{Gonzalez-Tudela2013} predict an ultimate upper limit of $\mathrm{g}^{(2)}(0) = 3$ reached at $\Omega_R = 150 \gamma$ and $\Gamma = 0.005 \gamma$. For solid-state emitters such as the QD studied here, this value may not be reached owing to phonon-mediated interactions that cause the coherent fraction to revive at large $\Omega_{R}$ \cite{PhysRevLett.110.217401}.

To further investigate filtering in the strong driving regime, Fig.$\:$\ref{fig:highpower}(c) presents a filter width dependence at constant $\Omega_{R} = 2\:\gamma$. At $\Gamma \gg \gamma$, antibunching is observed in accordance with the expectation for unfiltered RF. The antibunching in this region is degraded due to the period of the Rabi oscillations in $\mathrm{g}^{(2)}(t)$ (see  Fig.$\:$\ref{fig:highpower}(c) inset) being shorter than the detector IRF. Fig.$\:$\ref{fig:highpower}(d) shows the fraction ($\mathcal{F}$) of the filtered spectrum arising from each component for $\Omega_{R} = 2\:\gamma$. As $\Gamma$ becomes comparable to $\gamma$ in the central region of Fig.$\:$\ref{fig:highpower}(c), there is a transition to bunched photon statistics in accordance with Fig.$\:$\ref{fig:highpower}(a). This transition originates in the removal of the Mollow side peaks (green) from the filtered spectrum as $\Gamma$ decreases, combined with the onset of the indistinguishability bunching effect previously described.

As $\Gamma \ll \gamma$ is approached on the left-hand side of Fig.$\:$\ref{fig:highpower}(c), $\mathrm{g}^{(2)}(0)$ transitions again towards the  Poissonian statistics that were observed for $\Gamma \ll \gamma$ in Fig$\ldotp$ \ref{fig:filterwidth_timebroadening_g2}(a). The interpretation here is also the same; for such small $\Gamma$ the filtered spectrum contains almost solely coherent scattering (red line in \ref{fig:highpower}(d)) which exhibits Poissonian statistics when spectrally isolated. Ultimately, for very narrow filters of bandwidth comparable to the laser linewidth ($\sim 0.005 \gamma$), bunching would be expected to return due to indistinguishability bunching associated with the coherent part of the spectrum. Our theoretical model (green area in Fig.$\:$\ref{fig:highpower}(c)) successfully reproduces this behaviour, incorporating both the detector IRF and lower and upper bounds corresponding to the measured uncertainty ($0-20\%$) in the laser background contribution to the total signal (see Ref. \cite{sup}). It is interesting to note that the upper bound incorporating a $20 \%$ background exhibits stronger bunching than the lower bound, indicating the non-trivial effect of introducing an additional Poissonian background.

In summary, we have demonstrated that the resonance fluorescence spectrum of a two-level emitter comprises multiple interfering components that each exhibit distinct photon statistics. Without filtering, these components always interfere to produce the strong antibunching expected from single quantum emitters. However, when spectrally filtering with bandwidth comparable to the natural linewidth ($\gamma$) or Rabi frequency ($\Omega_{R}$), the ratio of these components is modified in the filtered spectrum, leading to strongly modified photon statistics. For weak resonant driving, a suitably narrow filter removes nearly the entire incoherent component, destroying the antibunching and illustrating that a sub-natural linewidth and strong antibunching cannot be simultaneously measured. For strong resonant driving, a pronounced bunching effect is observed at filter bandwidths comparable to the natural linewidth before the system ultimately trends towards Poissonian statistics for the narrowest filters. These results illustrate a potential new approach to manipulate the photon statistics of quantum light. In addition, we emphasise that care is required to preserve antibunching when filtering the spectrum of quantum emitters, an important consideration for future high throughput quantum networks where techniques such as wavelength-division multiplexing will be required.

\begin{acknowledgments}
This work was funded by the EPSRC (UK) EP/N031776/1, A.N. is supported by the EPSRC (UK) EP/N008154/1 and J.I.S. acknowledges support from the Royal Commission for the Exhibition of 1851. The authors would like to thank Andrew Foster and Dominic Hallett for assistance with the operation of the superconducting detectors, Tillmann Godde and Lisa Scaife for their work on laser power stabilisation and K\'{e}vin Seurre for contributing to the development of the tunable filter.
\end{acknowledgments}

\textit{Note added in proof.} -- Following the submission of our manuscript, we became aware of related results \cite{2020arXiv200511800H}.

\bibliography{Filtering}

\end{document}

\title{Supplemental Material: Photon Statistics of Filtered Resonance Fluorescence}

\author{Catherine L. Phillips}
\affiliation{Department of Physics and Astronomy, University of Sheffield, Sheffield, S3 7RH, United Kingdom}
\author{Alistair J. Brash}
\email[Email:]{a.brash@sheffield.ac.uk}
\affiliation{Department of Physics and Astronomy, University of Sheffield, Sheffield, S3 7RH, United Kingdom}
\author{Dara P. S. McCutcheon}
\affiliation{Quantum Engineering Technology Labs, H. H. Wills
Physics Laboratory and Department of Electrical and Electronic Engineering, University of Bristol, Bristol BS8 1FD, United Kingdom}
\author{Jake Iles-Smith}
\affiliation{Department of Physics and Astronomy, University of Sheffield, Sheffield, S3 7RH, United Kingdom}
\affiliation{Department of Physics and Astronomy, The University of Manchester, Oxford Road, Manchester M13 9PL, United Kingdom}
\affiliation{Department of Electrical and Electronic Engineering, The University of Manchester, Sackville Street Building, Manchester M1 3BB, United Kingdom}
\author{Edmund Clarke}
\affiliation{EPSRC National Epitaxy Facility, Department of Electronic and Electrical Engineering, University of Sheffield, Sheffield, UK}
\author{Benjamin Royall}
\affiliation{Department of Physics and Astronomy, University of Sheffield, Sheffield, S3 7RH, United Kingdom}
\author{Maurice S. Skolnick}
\affiliation{Department of Physics and Astronomy, University of Sheffield, Sheffield, S3 7RH, United Kingdom}
\author{A. Mark Fox}
\affiliation{Department of Physics and Astronomy, University of Sheffield, Sheffield, S3 7RH, United Kingdom}
\author{Ahsan Nazir}\email[Email:]{ahsan.nazir@manchester.ac.uk}
\affiliation{Department of Physics and Astronomy, The University of Manchester, Oxford Road, Manchester M13 9PL, United Kingdom}
\date{\today}
%\pacs{Valid PACS appear here}

\maketitle
\section{Theoretical Methods}
In this section we describe in detail the sensor theory used in the main text to fit the filtered resonance fluorescence data. We also discuss why phonon effects are expected to be largely absent from these experimental measurements.
\subsection*{The Sensor Formalism}
The sensor theory approach to calculating frequency-filtered $N$-photon correlation functions relies on enlarging the system Hilbert space to include $N$ auxiliary two-level systems that act as sensors of the emitted photons. %For two-time correlation functions, we require two auxiliary two-level systems. 
The formalism presented in this Supplement is based on Refs.~\cite{PhysRevLett.109.183601,PhysRevLett.116.249902}, extended to include the effects of laser background.

We consider a laser-driven quantum dot (QD) as a two-level system with ground-state $|g\rangle$ and excited state $|e\rangle$, split by energy $\epsilon$. In a frame rotating at the laser driving frequency and within the rotating-wave approximation on the driving, the QD Hamiltonian may be written (we take $\hbar=1$ throughout)
\begin{align}\label{hdot}
H_{\rm QD}=\nu|e\rangle\langle e|+\frac{\Omega_R}{2}\sigma_x,
\end{align}
where $\Omega_R$ is the Rabi frequency, $\nu=\epsilon-\omega_l$ is the QD-laser detuning, and $\sigma_x=|e\rangle\langle g|+|g\rangle\langle e|$. The driven QD is coupled to the electromagnetic field, which has the free Hamiltonian
\begin{align}
H_{\rm EM}=\sum_k\omega_ka_k^{\dagger}a_k.
\end{align}
The QD-field interaction, again in the rotating frame and within the rotating-wave approximation, is written
\begin{align}
H_{\rm QD-EM}=\sum_{k}(g_k\sigma^{\dagger}a_ke^{i\omega_lt}+g_k^*\sigma a_k^{\dagger}e^{-i\omega_lt}).
\end{align}
Here, $a_k$ annihilates a photon in mode $k$ of the electromagnetic field, which couples to the QD with strength $g_k$, and $\sigma=|g\rangle\langle e|$.

We may trace out the electromagnetic environment following the standard Born-Markov-secular approach assuming an electromagnetic field at zero temperature~\cite{carmichael1}. This allows us to derive an optical master equation governing the QD dynamics in the rotating frame:
\begin{align}
\dot{\rho}_{\rm QD}=-i[H_{\rm QD},\rho_{\rm QD}]+\gamma\mathcal{L}_\sigma(\rho_{\rm QD}).
\end{align}
Here, $\rho_{\rm QD}$ is the QD reduced density operator after tracing out the electromagnetic field, $\gamma=1/T_1$ is the QD spontaneous emission rate, and
\begin{align}\
\mathcal{L}_\alpha(O)=\alpha O\alpha^{\dagger}-\frac{1}{2}\{\alpha^{\dagger}\alpha,O\},
\end{align}
is the dissipator with anti-commutator $\{\cdot,\cdot\}$. We have neglected small energy shift terms that can anyway be absorbed into the definitions of $\nu$ and $\Omega_R$. The optical master equation is valid close to resonance, $\omega_l\sim\epsilon$, and for $\Omega_R\ll\epsilon$, both of which are fulfilled within the experiments.

To include the two-level sensors we enlarge our system Hilbert space, such that the system Hamiltonian now becomes (once more in the rotating frame and rotating-wave approximation)
\begin{align}
H_{\rm S}=H_{\rm QD}+\sum_{i=1}^{2}\nu_i\theta_i^{\dagger}\theta_i+\eta_i(\sigma\theta_i^{\dagger}+\sigma^{\dagger}\theta_i)+b_i\eta_i(\theta_i^{\dagger}+\theta_i),
\end{align}
where $\theta_i=|g_i\rangle\langle e_i|$ is the lowering operator for sensor $i$, which is centred at (and thus filters around) frequency $\omega_i$, with $\nu_i=\omega_i-\omega_l$. The QD-sensor coupling is given by $\eta_i$, which will eventually be taken to be vanishingly small. We also include laser background terms with strengths $b_i$, modelled phenomenologically as direct excitation of the sensors. These terms are necessary due to the imperfect polarisation rejection of the excitation laser within the experiments. We consider two sensor systems, as our primary interest is in second-order photon correlations.

Given that the QD-sensor couplings are vanishingly small, the form of the QD emission dissipator $\mathcal{L}_{\sigma}$ is unaffected by the presence of the sensor systems. Likewise, broadening of the sensors, and hence the filter width, can be described by dissipators acting individually on each sensor system, such that our master equation now becomes~\cite{PhysRevLett.109.183601,PhysRevLett.116.249902}
\begin{align}
\dot{\rho}_{\rm S}=-i[H_{\rm S},\rho_{\rm S}]+\gamma\mathcal{L}_\sigma(\rho_{\rm S})+\Gamma_1\mathcal{L}_{\theta_1}(\rho_S)+\Gamma_2\mathcal{L}_{\theta_2}(\rho_S),
\end{align}
where $\Gamma_i$ is the filter width for sensor $i$. 
The frequency- and time-resolved two-photon correlation function is given by~\cite{PhysRevLett.116.249902}
\begin{align}
g^{(2)}_{\Gamma_1,\Gamma_2}(\nu_1,T_1;\nu_2,T_2)=\lim_{\eta_1,\eta_2 \to 0}\frac{\langle:n_1(T_1)n_2(T_2):\rangle_{\mathcal T}}{\langle n_1(T_1)\rangle\langle n_2(T_2)\rangle},
\end{align}
where $n_i(t)=\theta^{\dagger}_i(t)\theta_i(t)$, and $\langle:\;:\rangle_{\mathcal T}$ indicates normal and time ordering.

The experiments are performed under resonant continuous driving conditions, such that $\nu=0$ in Eq.~\ref{hdot}, with the filters centred on the QD transition, meaning that $\nu_1=0=\nu_2$ as well. In the steady-state, the second-order photon correlation function then becomes
\begin{align}
g^{(2)}_{\Gamma}(\tau)=\lim_{\eta \to 0}\frac{{\rm tr}[n_2(\tau) \theta_1(0)\rho_{\rm S}(\infty)\theta_1^{\dagger}(0)]}{\langle n_1(0)\rangle_{\rm ss}\langle n_2(0)\rangle_{\rm ss}},
\end{align}
where $\rho_{\rm S}(\infty)$ is the steady-state density operator of the combined system of the QD and sensors, $\langle n_i(0)\rangle_{\rm ss}={\rm tr}[n_i(0)\rho_{\rm S}(\infty)]$, we set $\Gamma_1=\Gamma_2=\Gamma$ as the filter width, and $\eta_1=\eta_2=\eta$. Multi-time correlation functions are calculated using regression in the standard way~\cite{carmichael1}.

The parameters $\Omega_R$, $\gamma$, and $\Gamma$ are known in the experiments, and so take fixed values within the theoretical calculations. The only free parameter is the background level, $b_1=b_2=b$, which is used as a fitting parameter (constrained by experimental measurements - see subsequent section) for the insets of Fig.~4 (a) and (c), as well as to give the confidence bounds in the main panel of Fig.~4 (c). When considering the detector instrument response we convolve $g^{(2)}_{\Gamma}(\tau)$ with the Gaussian function
\begin{align}
\mathcal{I}(t)=\frac{2}{\delta t}\sqrt{\frac{\ln{2}}{\pi}}e^{-4\ln{2} (t/\delta t)^2},
\end{align}
where $\delta t$ gives the full-width half-maximum.

\begin{figure}[b]
	\includegraphics[width=\columnwidth]{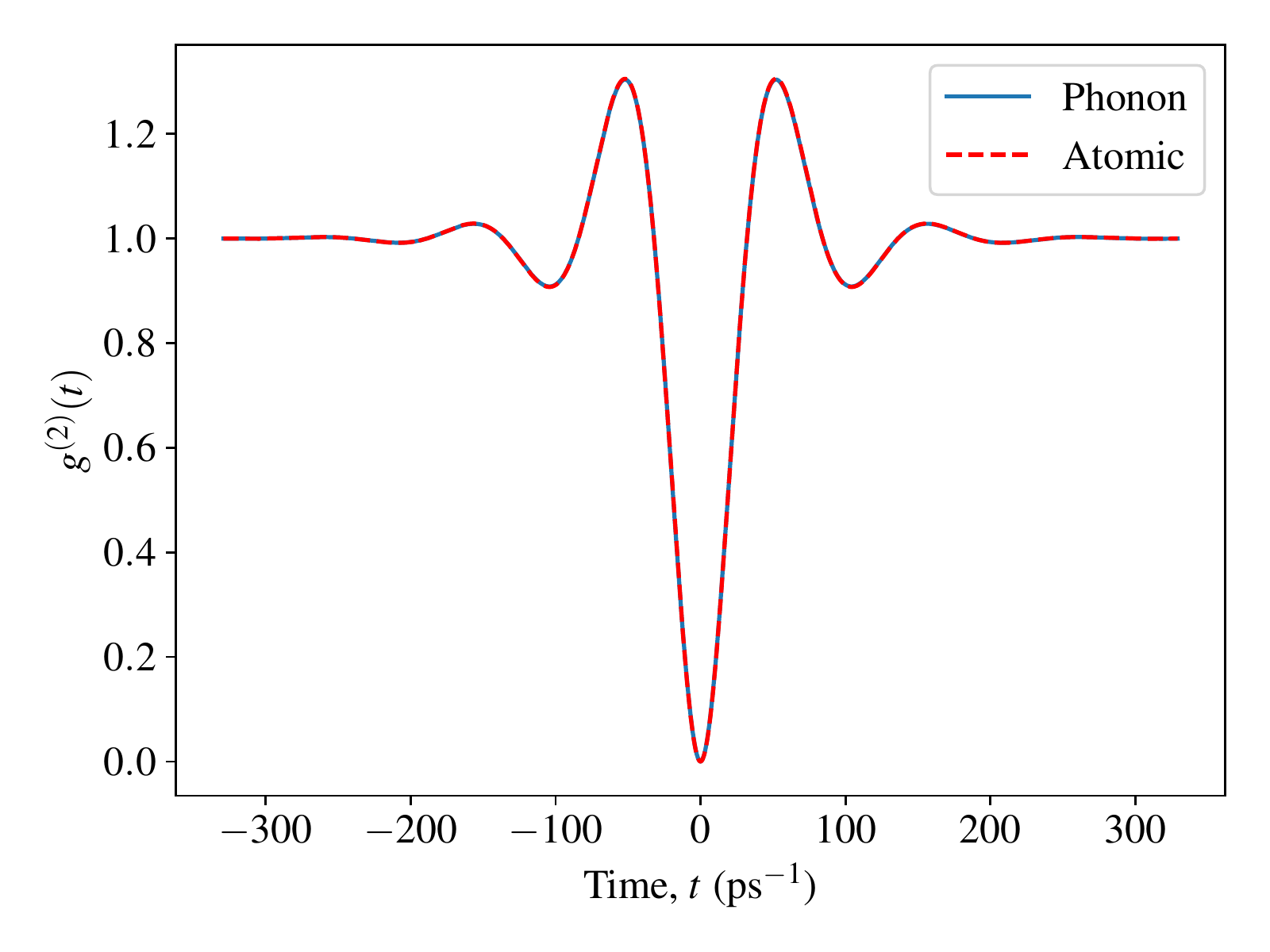}
	\caption{The second order correlation function calculated including phonons through a polaron theory approach (solid), and a simplified atomic theory in the absence of phonons (dashed). In keeping with the measured sample, the spontaneous emission rate is $\gamma = 20~\mu$eV and the renormalised Rabi frequency is $\Omega_R = 2\gamma$. 
	The phonon parameters were found in Ref.~\cite{PhysRevLett.123.167403} through fitting to the emission spectra of the same QD used in the present experiments. The temperature is set to $4$~K.} 
\end{figure}

\subsection*{Electron-phonon coupling and $g^{(2)}(t)$}
Usually, phonon coupling plays a prominent role in determining the characteristics of QD emission and the associated photon correlation functions~\cite{PhysRevB.95.201305,PhysRevLett.123.167403}. 
This is through three mechanisms, Rabi frequency renormalisation~\cite{PhysRevLett.105.177402}, excitation induced dephasing~\cite{PhysRevLett.104.017402}, and phonon sideband emission~\cite{PhysRevB.95.201305,PhysRevLett.123.167403}.

In the first of these mechanisms, the dipole of the quantum emitter is dressed by modes of the phonon environment~\cite{PhysRevLett.105.177402}. 
This results in the effective Rabi frequency being reduced from the value expected in the absence of phonons. 
It is thus this renormalised Rabi frequency which is observed in the experiments presented in the main manuscript, and so the renormalisation effect is already implictly included in the model by our use of the observed Rabi frequencies in the theoretical calculations. 

In contrast, excitation induced dephasing has negligible impact on the measured values of $g^{(2)}$. This is due to the nature of such correlation measurements; the second order optical coherence of a quantum emitter principally probes the level structure of the emitter, and is naturally insensitive to coherence and dephasing. 
To emphasise this point, Fig. S1 compares $g^{(2)}$ as a function of time in the absence of filtering as calculated using a full phonon theory based on the polaron formalism~\cite{Nazir_2016}, and a simple atomic theory using the renormalised Rabi frequency as outlined above.
As can be seen, the resultant dynamics are identical, suggesting that excitation-induced phonon effects are negligible for the quantities of interest in this work.
%Note that for consistency, the renormalised Rabi frequency is used in the atomic theory as the is the Rabi frequency that would be observed in experiment.

Finally, a phonon sideband is also present in the emission of the QD~\cite{PhysRevB.95.201305,PhysRevLett.123.167403}.
From the measured data in the regions where $\Gamma > 10 \:\gamma$ in Figs. 2(a) and 4(c), it is clear that filtering the sideband has no impact on the photon statistics. 
We interpret this as being a consequence of the nature of the sideband. Each photon emitted through the sideband is naturally correlated with a phonon, with the state becoming mixed when one traces out the phonon degrees of freedom. This prevents interference between the sideband and the zero-phonon line scattering of the quantum emitter, which would be necessary to see changes in the photon statistics.

\section{Experimental Methods}

In this section we provide further detail on the experimental methods used including the filtering set-up, experimental spectrum and signal to background ratio.

\subsection{Spectral Filters}
\begin{table}[h]
	\centering
	\begin{tabular}{|c|c|}
		\hline
		\textbf{Filter} & \textbf{Bandwidth} $\Gamma$ ($\upmu$eV)\\ \hline\hline
		Free-space Fabry-P\'{e}rot & 0.25\\ \hline
		Etalon - 1.6mm fused silica & 5.8\\ \hline
		Fibre Fabry-P\'{e}rot & 17\\ \hline
		1200 l mm$^{-1}$ grating spectrometer & 97 \\ \hline
		4f tunable filter (narrow) & 454\\\hline
		4f tunable filter (broad) & 3050\\ \hline
	\end{tabular}
	\caption{Bandwidth ($\Gamma$) of spectral filters.}
	\label{table:filterbandwidth}
\end{table}
Table~\ref{table:filterbandwidth} lists the filters used in these measurements. The filter with the greatest bandwidth ($\Gamma$) uses a folded-geometry 4f pulse shaper design with a 1500~l~mm$^{-1}$ transmission grating. This design allows tuning of $\Gamma$ over a range of $>6$~meV by altering the width of an adjustable slit in the focal plane of the pulse shaper. The $\Gamma$ of the filter as a function of slit width was measured with a broadband white-light source; the transmitted spectrum was recorded on a spectrometer.

The spectrometer itself ($\mathrm{f}=0.75~\mathrm{m}$, 1200~l~mm$^{-1}$) can also be used as a spectral filter; a diverter mirror in front of the camera instead directs the frequency separated emission through an exit slit. After the slit, the emission is recoupled to a single mode fibre. The $\Gamma$ of the spectrometer filter depends on the grating used as well as the exit slit width and coupling back into single mode fibre. $\Gamma$ was measured by using the spectrometer to filter a broadband attenuated laser pulse from a femtosecond Ti:S laser. The width of the filtered pulse was measured on a second spectrometer and found to be 97~$\upmu$eV.

Both of the ``Fabry-P\'{e}rot" filters are scanning interferometers (FPIs). Scanning interferometers use highly reflective mirrors to create a cavity where constructive intereference at the cavity boundary allows the transmission of a small wavelength range. The central wavelength of this transmission can be tuned by changing the length of the cavity. To achieve this, one of the mirrors is mounted on a piezo so that the cavity length can be tuned by applying a voltage. To use a scanning interferometer as a narrow bandwidth filter, the transmission wavelength is locked by continuously adjusting the applied voltage to negate any thermal drifts in cavity length. In this work the filtering is centred on the ZPL, thus the central transmission wavelength of the Fabry-P\'{e}rot is locked by continuously maximising the intensity recorded on a single photon detector. $\Gamma$ of the Fabry-P\'{e}rot filters was found using the following equation:
\begin{equation}
    \Gamma = \frac{FSR}{\mathcal{F}},
\end{equation}
where $FSR$ is the free spectral range and $\mathcal{F}$ is the finesse of the cavity. Both of these values were supplied in the manufacturer's test reports.

The Etalon filter works similarly to the Fabry-P\'{e}rot filters; cavity interference effects lead to transmission of only a small bandwidth. However, the Etalon is solid fused silica with reflective coatings rather than a tunable mirror separation design. The centre wavelength of the etalon can instead be tuned by tilting the etalon with respect to the incoming beam. The $\Gamma$ of the Etalon filter was ascertained by measuring the transmission of a tunable CW Ti:S laser on a power meter, the resulting Lorentzian transmission spectrum gave $\Gamma =$ 5.8~$\upmu$eV.

\subsection{Spectrum}

\begin{figure}[h!]
    \centering
    \includegraphics{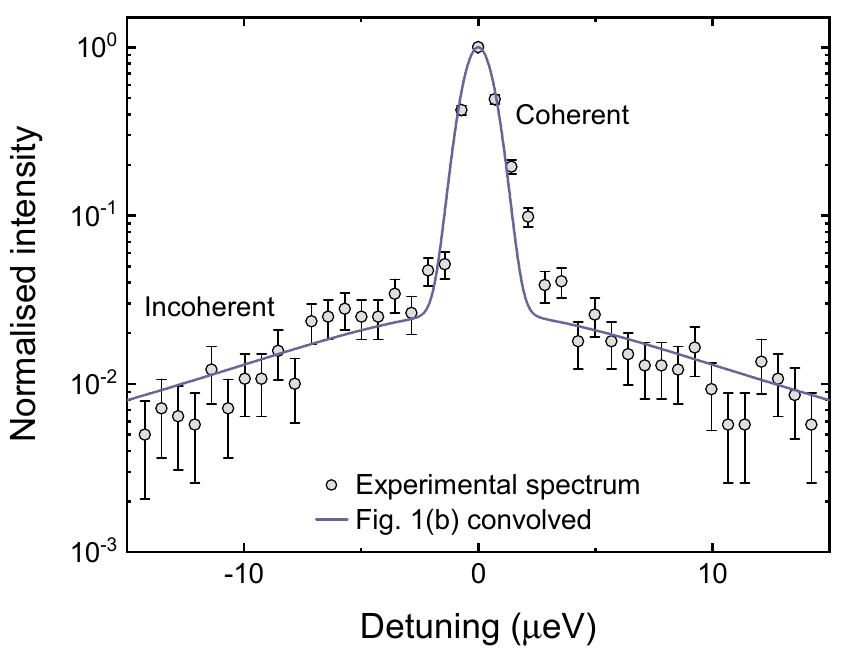}
    \caption{High resolution spectrum of the QD emission measured with a scanning Fabry-P\'{e}rot interferometer (FPI) at $\Omega_R = 0.5\gamma$. Grey circles - experimental spectrum, purple line - theoretical spectrum (see Fig. 1(b)) after convolution with the FPI IRF (Gaussian, $1.5~\upmu \mathrm{eV}$ FWHM.)}
    \label{fig:ExpSpec}
\end{figure}

Fig. \ref{fig:ExpSpec} shows a high resolution spectrum measured by sweeping the free-space FPI detailed in the previous section and recording the transmitted intensity measured on a single photon detector. As the spectrum is acquired over many sweeps, thermal drifts give a Gaussian IRF of $1.5~\upmu \mathrm{eV}$ when the instrument is operated in this configuration. The grey circles in Fig. \ref{fig:ExpSpec} show the experimental result which comprises a relatively narrow coherent peak and a much broader incoherent peak as expected from Fig. 1(b). Convolving the theoretical spectrum of Fig. 1(b) with the IRF of the scanning FPI significantly broadens the coherent peak, resulting in the purple curve in Fig. \ref{fig:ExpSpec} that agrees well with the experimental data.

\subsection{Signal to Background}

\begin{figure}[h]
    \centering
    \includegraphics{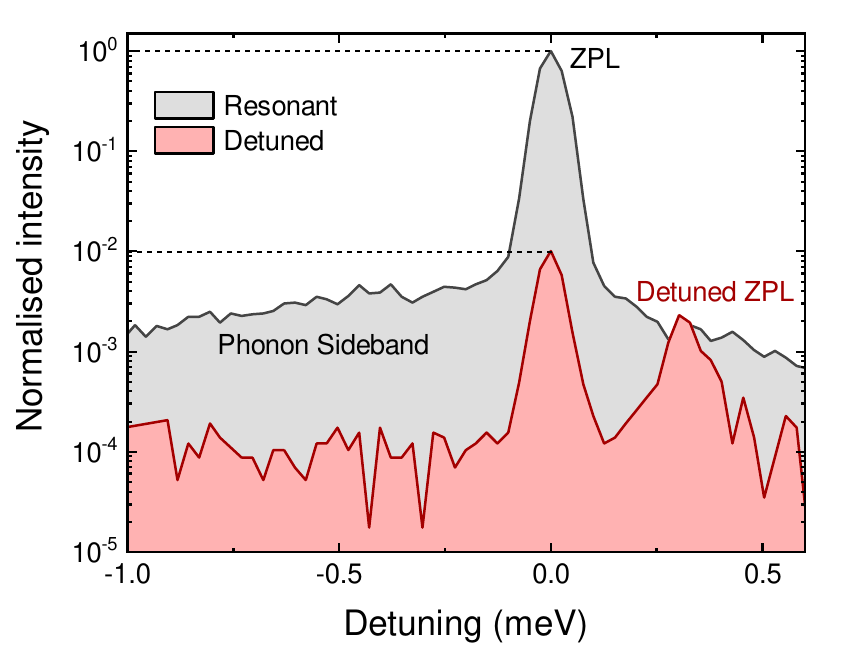}
    \caption{Spectrum of the QD emission taken at $\Omega_{R} = 0.5 \gamma$; grey line -  ZPL resonant with laser, red line - ZPL electrically detuned by +0.3~meV from the laser. Comparison of the area of the zero detuning peaks allows us to place a lower bound on the signal to background ratio of 100:1 (see dashed lines). For the resonant case, a broad but weak phonon sideband is also observed.}
    \label{fig:OnOffSpec}
\end{figure}

As the sample incorporates a p-i-n diode structure for electrical tuning of the QD transitions, the signal to background ratio (SBR) of our resonance fluorescence measurements can readily be checked by tuning the transition out of resonance with the laser. An example of such a measurement is presented in Fig. \ref{fig:OnOffSpec} for $\Omega_R = 0.5 \gamma$; comparison of the areas of the zero detuning peaks gives a SBR of 100:1, demonstrating conclusively that the observed behaviour originates from the QD signal. This value represents a lower bound as the laser peak for the detuned case also contains some coherent scattering from the detuned emitter. This level of laser background ($< 1\%$) has a negligible effect on our $\mathrm{g^{(2)}(t)}$ measurements and is thus disregarded in the theoretical calculations for Fig. 2 in the main manuscript.

Applying the same methodology at $\Omega_{R} = 2 \gamma$ gives a SBR of around 10:1, this degradation compared to $\Omega_{R} = 0.5 \gamma$ is due to the 16-fold increase in laser power whilst the RF signal itself is beginning to saturate. Over the course of a $\mathrm{g^{(2)}(t)}$ measurement, thermal drifts of polarisation optics in our set-up can degrade this to a measured SBR of 4:1 by the end of the measurement. As these levels of background can be significant, they are included in our theoretical model as the parameter $b$ (see prior section on the sensor formalism). Our experimental measurements of SBR are used as constraints when fitting the $\mathrm{g^{(2)}(t)}$ data in the insets of Figs. 4(a) and 4(c). The observed uncertainty in SBR during the $\mathrm{g^{(2)}(t)}$ measurement also gives rise to the confidence bounds in the main plot of Fig. 4(c) with the maximum background level of $20\%$ corresponding to the worst case of 4:1 SBR.

%\bibliography{supp_refs}% Produces the bibliography via BibTeX.
\makeatletter\@input{xx.tex}\makeatother